\documentclass[aip,jcp,reprint]{revtex4-1}
\usepackage{braket}
\usepackage{amsmath}
\usepackage{xfrac}
\usepackage{hyperref}
\renewcommand{\thispagestyle}[1]{}

\begin{document}
\title{Efficient Simulation of Near-Edge X-ray Absorption Fine Structure (NEXAFS) in
Density-Functional Theory: Comparison of Core-Level Constraining Approaches}
\author{Georg S. Michelitsch}
\affiliation{Chair for Theoretical Chemistry and Catalysis Research Center, Technische Universit{\"a}t M{\"u}nchen, Lichtenbergstr. 4, D-85748 Garching, Germany}
\email[]{georg.michelitsch@ch.tum.de}
\author{Karsten Reuter}
\affiliation{Chair for Theoretical Chemistry and Catalysis Research Center, Technische Universit{\"a}t M{\"u}nchen, Lichtenbergstr. 4, D-85748 Garching, Germany}
\date{\today}

\begin{abstract}
	Widely employed Near-Edge X-Ray Absorption Fine Structure (NEXAFS) 	spectroscopy probes a system by excitation of core electrons to unoccupied states. A variety of different methodologies are available to simulate corresponding spectra from first-principles. Core-level occupation constraints within ground-state Density-Functional Theory 	(DFT) represent a numerically most efficient means to this end that provides access to large systems, examples being surface adsorption, 	proteins, polymers, liquids, and buried, condensed phase interfaces 	(e.q. solid-liquid and solid-solid). Here, we systematically investigate the performance of different realizations of this approximate approach through the simulation of K-edge 	NEXAFS-spectra of a set of carbon and nitrogen-containing organic molecules. Variational collapse to the ground state and oscillatory convergence are the major complications of these approximate computational protocols. We present a modified version of the maximum-overlap method to achieve a self-consistent inclusion of electrons in virtual states for systems where convergence is hampered due to degeneracies. Our results demonstrate that reliable spectra allowing for a semi-quantitative analysis of experimental data are already obtained at the semi-local level of density functionals and with standard numeric atomic orbital basis sets.
\end{abstract}


\maketitle 

\section{Introduction}

Core-level spectroscopies are among the most established characterization techniques in modern materials science, providing both chemical and structural information. In modern nanosciences, not only X-Ray Photoelectron Spectroscopy (XPS), but also advanced techniques such as Near-Edge X-Ray Absorption Fine Structure (NEXAFS) are important tools to study molecules in the gas phase as well as molecules or thin layers of molecules immobilized on a support.\cite{Stohr1992,Chen1997,Kolczewski2001,Wende2004,Rehr2005,Ostrom2007a,Fronzoni2014,Gattinoni2019}
Furthermore, dynamical systems such as liquids\cite{Kong2012,Wernet2004,Iannuzzi2008,Chen2010,Hetenyi2004,Smith2017,Matsui2016}
and soft matter such as molecular crystals\cite{Schwartz2010}, polymers\cite{Perera2018,Su2017}, and proteins\cite{Stewart-Ornstein2007} are objects of intense study, followed by the dynamics of material growth as for example graphene on copper\cite{Rojas2018} and oxidation processes of bulk condensed matter\cite{Poloni2018}. Always exciting a (core) electron in an energetically low-lying state through X-ray radiation, it is the energy and type of radiation that distinguishes different such spectroscopies.
In XPS, the electron is entirely ejected, whereas in NEXAFS and related techniques the electron is excited to an unoccupied state. More information about the system can be obtained by multiple measurements with different polarity of the light (revealing magnetic properties) or at varying incidence angles (revealing orientational differences). 

Notwithstanding this versatility, in surface-adsorption, supramolecular or dynamically changing systems, the multiplicity of chemical environments for the same species renders a straightforward interpretation of experimental spectra increasingly complex.
Many overlapping peaks in the same energetic region combine to a single unresolved and broadened peak, while orbital hybridization diminishes the usefulness of reference spectra obtained for gas-phase molecules. In this situation, simulated spectra from independent first-principles calculations become invaluable for a reliable assignment.\cite{Diller2017,Kolczewski2007,Puttner2017,Cavalleri2009,Zhovtobriukh2018,Fransson2016,Su2017,Drisdell2017,Hellgren2001,Gao2009a}
However, especially for large systems such as frequently encountered in supramolecular or surface-adsorption contexts exceeding computational costs largely restrict the types of methodology that can be employed. While in principle highly accurate techniques such as time-dependent density-functional theory (DFT) \cite{Stener2003,Besley2010,Verma2016a,Nakata2007,Besley2009a,Nakata2006}, the Bethe-Salpeter approximation\cite{Vinson2012,Vinson2011}, coupled-cluster approaches\cite{Nooijen1995,Coriani2012,Peng2015}, or multi-reference calculations\cite{Coe2015,Bokarev2013,Grell2015} are available for an often quantitative simulation of NEXAFS spectra, in practice it is presently often only effective core-level occupation constraining approaches in ground-state DFT that are numerically feasible. This is especially true in cases of dynamically changing systems, where the experimental signature is a combination of many different molecular arrangements (e.q. liquids\cite{Kong2012,Wernet2004,Iannuzzi2008,Chen2010,Hetenyi2004,Smith2017,Matsui2016}) or a large number of possible (yet chemically different) excitation centers (e.g.
proteins\cite{Stewart-Ornstein2007}).
In these aforementioned effective constraining approaches, specific occupations of single-particle Kohn-Sham (KS) levels are enforced to mimic the core-excited state, and then the lowest-energy electronic configuration under this constraint is self-consistently determined.\cite{Egelhoff1987,Triguero1999,Takahata2003,Matteo2005,Zeng2010,Garcia-Gil2012}
On the positive side, this captures a dominant contribution to the important core-hole relaxation energy at numerical costs that are at the level of a regular ground-state DFT calculation. On the negative side, different ways of changing the occupation of the targeted core and virtual states give rise to a range of differing computational protocols in this class of techniques. Most importantly, there are variants that explicitly consider the occupation of the formerly unoccupied KS state, requiring multiple calculations for different final states to assemble the total NEXAFS spectrum.\cite{Bagus1965,Slater1972}
Other so-called implicit variants such as the Transition Potential (TP)\cite{Triguero1998} and eXcited electron and Core Hole (XCH)\cite{Prendergast2006} method either neglect the excited final-state electron or only include it in an averaged way, and would, therefore, allow to compute a full spectrum with only one single calculation. As such the computational effort to simulate the spectroscopic signature can vary largely between different variants, while the advantage in terms of accuracy is often not clear. Although by explicitly considering the excited electron in the simulation better results are expected, the realization of such simulations is often impossible in practice due to problems associated with variational collapse and convergence of the electronic structure. Here, we partially address this problem with the introduction of a variant of the maximum overlap method\cite{Besley2009}, optimized for the usage in highly symmetric systems plagued by degeneracies. As has also been noted earlier\cite{Uejio2008}, local basis set based approaches typically have problems to converge resonances above the ionization threshold. We acknowledge this problem (and further details on the performance of the MOM approach in our case can be seen in the supporting information) and also recognize it as probably one of the major arguments why we are interested in implicit variants, which by construction, eliminate the need for the inclusion of an excited electron.  As for the variety of these implicit variants, we wish to elaborate on the different motivations as of why they were introduced and classify them according to similarities. This should help in the understanding as to which variant should be chosen based on the system under study and as of how the accuracy can be systematically improved. While establishing this hierarchy, we noticed the presence of gaps in terms of implicit approximations. We filled this gaps with the introduction of the Generalized Transition Potential (GTP) and eXcited transition potential (XTP), as well as eXcited Generalized Transition Potential (XGTP) approaches. In the current manuscript, we do not include mixed approaches which either correct selected excitation energies of an implicit spectrum by the explicit calculation via a $\Delta$SCF ansatz\cite{Kolczewski2001} or via explicit modeling of the chemical shift of each atom via an additional explicit consideration of the lowest possible transition\cite{Nyberg1999}.

As particularly the class of explicit variants requires an adequate description of the (typically more diffuse) unoccupied KS states, a number of studies have assessed the numerical convergence of correspondingly simulated spectra for more common localized (Gaussian) basis sets.\cite{Fouda2018,Tolbatov2014,Takahata2000,Carniato2002,Chong2006,Takahata2003,Chong2002,Besley2009}
In contrast, much less is known on the basis set requirements of the latter class of implicit variants (intuitively deemed less demanding) and generally for numeric atomic orbital (NAO) type basis sets.\cite{Blum2009} Aiming to establish a numerically most efficient, yet robust protocol for large-scale NEXAFS simulations with NAO basis sets as for instance implemented in the full-potential DFT code FHI-aims \cite{Blum2009,Ren2012}, we, therefore, present a systematic investigation using a test set of nitrogen- and carbon-containing compounds. With an eye to maximally support the experimental assignment, we evaluate the influence of different basis sets and DFT functionals on both the correct peak positions and the peak intensities. We include variants explicitly treating the final state, like $\Delta$ Self-Consistent-Field ($\Delta$SCF)\cite{Bagus1965} or the Transition State (TS)\cite{Slater1972}
model, and more approximate implicit variants like TP \cite{Triguero1998} or XCH \cite{Prendergast2006,Lie2000,Buczko2000,Ahuja1996}. The major and encouraging
result is that a semi-quantitative spectral assignment is already possible for numerically most efficient implicit variants, standard basis set sizes and semi-local DFT functionals.

\section{Theory}

\subsection{Core-hole constraining approaches}\label{sec:constraints}
\begin{figure}[htbp!]
	\setlength{\tabcolsep}{0.0em} 
	\begin{tabular}{cc|ccc|ccc|ccc}
		\multicolumn{2}{c|}{} & 
		\multicolumn{3}{c|}{explicit} & 
		\multicolumn{3}{c|}{implicit} & 
		\multicolumn{3}{c}{neutral implicit} \\[0.2em] 
		\hline
		&
	\scriptsize	\rotatebox{50}{GS} &
	\scriptsize	\rotatebox{50}{$\Delta$SCF$^{\ddagger}$} &
	\scriptsize	\rotatebox{50}{TS$^{\ddagger}$} &
	\scriptsize	\rotatebox{50}{GTS$^{\ddagger}$} &
	\scriptsize	\rotatebox{50}{FCH} &
	\scriptsize	\rotatebox{50}{TP} &
	\scriptsize	\rotatebox{50}{GTP} &
	\scriptsize	\rotatebox{50}{XCH} &
	\scriptsize	\rotatebox{50}{XTP} &
	\scriptsize	\rotatebox{50}{XGTP}\\
	& 
	\multicolumn{1}{c}{\includegraphics[width=0.046\textwidth]{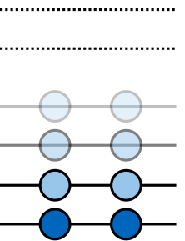}} &
	\multicolumn{1}{c}{\includegraphics[width=0.046\textwidth]{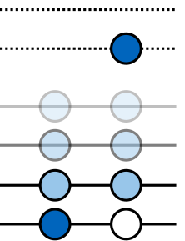}} &
	\multicolumn{1}{c}{\includegraphics[width=0.046\textwidth]{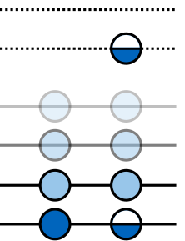}} &
	\multicolumn{1}{c}{\includegraphics[width=0.046\textwidth]{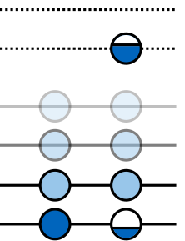}} &
	\multicolumn{1}{c}{\includegraphics[width=0.046\textwidth]{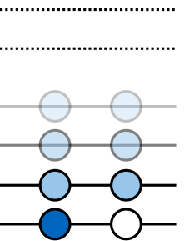}} &
	\multicolumn{1}{c}{\includegraphics[width=0.046\textwidth]{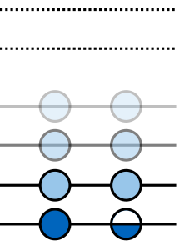}} &
	\multicolumn{1}{c}{\includegraphics[width=0.046\textwidth]{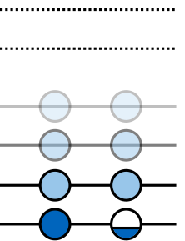}} &
	\multicolumn{1}{c}{\includegraphics[width=0.046\textwidth]{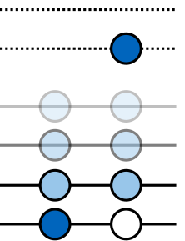}} &
	\multicolumn{1}{c}{\includegraphics[width=0.046\textwidth]{./TS.eps}} &
	\multicolumn{1}{c}{\includegraphics[width=0.046\textwidth]{./GTS.eps}} \\
	&
	$\Delta\epsilon$ &
	$\Delta E$ &
	$\Delta\epsilon$ &
	$\Delta\epsilon$ &
	$\Delta\epsilon$ &
	$\Delta\epsilon$ &
	$\Delta\epsilon$ &
	$\Delta\epsilon$ &
	$\Delta\epsilon$ &
	$\Delta\epsilon$ \\
	\hline
	$q_{\rm c}$ & 1 & 0 & \sfrac{1}{2} & \sfrac{1}{3} & 0 & \sfrac{1}{2} & \sfrac{1}{3} & 0  & \sfrac{1}{2} &\sfrac{1}{3} \\
	$q_{\rm v}$ & 0 & 1 & \sfrac{1}{2} & \sfrac{2}{3} & 0 & 0 & 0 & 1 & \sfrac{1}{2} & \sfrac{2}{3} \\
	\end{tabular}
	\caption{Schematic illustration of the occupational constraints used in various core-hole constraining approaches. The 
	table shows the corresponding fractional occupations of the core,
	$q_{\rm c}$, and virtual, $q_{\rm v}$, Kohn-Sham 
	eigenstate, whether the evaluation is based on total energy differences ($\Delta E$) or KS eigenvalue differences ($\Delta 
	\epsilon$), and whether the scheme explicitly considers the final-state
excited electron ($\ddagger$) or not. Explicit consideration requires that one
separate calculation needs to be performed for each excited state configuration
on each atom (many $q_{\rm v}$ for a single $q_{\rm c}$), whereas the implicit neutral approaches typically explore only the
lowest energy core-excited state (a single $q_{\rm c}$ constraint per atom).}
	\label{fig:approximations}
\end{figure}
Core-level occupation constraining approaches generally rely on time-dependent perturbation theory to compute the NEXAFS spectrum using Fermi's golden rule \begin{equation}
P_{{\rm i}\rightarrow {\rm f}}(\omega) \;=\; \frac{2\pi}{\hbar} {\bf \mu}_{\rm if}^2 \Delta(E_{\rm f} - E_{\rm i} - \hbar \omega) \quad .
\label{eq1}
\end{equation}
An incident X-ray with frequency $\omega$ induces an electronic transition from an initial state i to a final state f with matching energy difference $\Delta E = E_{\rm f} - E_{\rm i}$ with a probability proportional to the transition dipole moment ${\bf \mu}_{\rm if}^2$. In order to determine this probability within ground-state DFT the excited-state energy $E_{\rm f}$ is then approximately computed by modifying the occupation of the single-particle KS states and achieving self-consistency under this occupational constraint.
Various variants differ in the way how these occupations are modified, and whether they explicitly optimize every transition ${\rm i} \rightarrow {\rm f}$ separately or do this only implicitly in an average way. They are graphically summarized in Figure \ref{fig:approximations} and will be shortly introduced in the following.

In $\Delta$SCF \cite{Bagus1965} the excited-state energy is computed as a total energy difference by explicitly removing one electron from the corresponding core level c and adding it to the virtual level v, resulting in a transition energy \begin{align}
\nonumber \Delta E_{\Delta{\rm SCF}} = E_{\rm f} - E_{\rm i} = E(q_{\rm c} = 0,q_{\rm v} = 1) - \\ 
  			  	         E(q_{\rm c} = 1, q_{\rm v} = 0) \quad . \label{eqn:dscf} 
\end{align}
Here, $q_{\rm c}$ is the occupation of the core-state KS orbital and $q_{\rm v}$ is the occupation of the virtual KS state above the Fermi level. Throughout the work, we thereby stay within the realm of collinear spin-resolved DFT, where the maximum occupancy of a KS orbital is 1, and we follow the convention to denote total energies with negative numbers; the more negative, the more stable. 

Other core-hole constraining approaches use this basic equation of $\Delta$SCF as the starting point, rewrite it as an integral over the varying occupations during the electronic transition and employ the Slater-Janak theorem\cite{Janak1978} $\frac{\partial E}{\partial q_{\rm i}} = \epsilon_{\rm i}$ to arrive at eq. (\ref{eqn:dscf}) in terms of KS eigenvalues $\epsilon_{\rm i}$:
\begin{align}
\nonumber \Delta E_{\Delta{\rm SCF}} & =\\
\nonumber & = \int_{x=1}^0 \frac{dE(q_{\rm c} = x, q_{\rm v} = 1-x)}{dx} dx \\
\nonumber & = \int_{x=1}^0 \left\{ \frac{\partial E(q_{\rm c} = x, q_{\rm v} = 1-x)}{\partial q_{\rm v}}  \right. \\
\nonumber & \quad \quad \quad \quad - \left. \frac{\partial E(q_{\rm c} = x, q_{\rm v} = 1-x)}{\partial q_{\rm v}} \right\} dx \\
\nonumber & = \int_{x=1}^0 \left\{ \epsilon_{\rm c}(q_{\rm c} = x, q_{\rm v} = 1-x) \right. \\
					& \quad \quad \quad \quad - \left. \epsilon_{\rm v}(q_{\rm c} = x, q_{\rm v} = 1-x) \right\} dx \quad .
\label{eqn:dscf2}
\end{align}
Here, we performed a substitution and split the integral in two parts, because $\frac{\partial q_c}{\partial x} = 1$ and $\frac{\partial q_v}{\partial x} = -1$. In Slater's Transition State (TS) approach \cite{Slater1972}, the integral in eq. (\ref{eqn:dscf2}) is approximated via the midpoint rule $\int_a^b f(x) dx \simeq (b-a)f((a+b)/2)$. This results in \begin{eqnarray}
\Delta E_{\rm TS} = \epsilon_{\rm v}(q_{\rm c} = 0.5, q_{\rm v} = 0.5) - \\
\nonumber\epsilon_{\rm c}(q_{\rm c} = 0.5, q_{\rm v} = 0.5) \quad .
\end{eqnarray}
and bears the advantage that the transition energy can be obtained from two KS levels of one constrained-occupation DFT calculation. The Generalized Transition State (GTS) variant instead approximates the integral of eq. (\ref{eqn:dscf2}) by a two-point Gaussian quadrature including the ground state ($x=0$) as the first point and $x = 1/3$ as the second, thereby lowering the integration error to fourth order \cite{Williams1975,Chong1995a}
\begin{align}
\nonumber \Delta E_{\rm GTS} = & \\
\nonumber\left[\frac{1}{4}\epsilon_{\rm v}(q_{\rm c}=1, q_{\rm v} = 0) + \frac{3}{4}\epsilon_{\rm v}(q_{\rm c}=\sfrac{1}{3}, q_{\rm v} =\sfrac{2}{3})\right] & - \\
\left[\frac{1}{4}\epsilon_{\rm c}(q_{\rm c}=1, q_{\rm v} = 0) +
\frac{3}{4}\epsilon_{\rm c}(q_{\rm c}=\sfrac{1}{3}, q_{\rm v}
=\sfrac{2}{3})\right] &\quad .\label{eqn:gts}
\end{align}

$\Delta$SCF, TS and GTS all consider explicitly into which virtual state v the core electron is excited to. These explicit core-hole constraining variants therefore require a separate calculation for every transition i$\rightarrow$f to assemble the full NEXAFS spectrum. Implicit variants instead deem the actual impact of the excited electron on the KS level positions less important. Several of these variants therefore modify the occupation of the core level c, but leave the virtual level v indeed unoccupied also in the approximate calculation of the final-state energy $E_{\rm f}$. These variants include the Transition Potential (TP) \cite{Triguero1998,Stener1995,Hu1996} and the Generalized Transition Potential (GTP) variant, representing the direct implicit analogs to TS and GTS:
\begin{align}
	\nonumber \Delta E_{\rm TP} = & \epsilon_{\rm v}(q_{\rm c}=0.5, q_{\rm v} = 0) - \\
				  & \epsilon_{\rm c}(q_{\rm c}=0.5, q_{\rm v} =
	0)\quad .\label{eqn:tp}
\end{align}
\begin{align}
	\nonumber \Delta E_{\rm GTP} = & \\
 \nonumber\left[\frac{1}{4}\epsilon_{\rm v}(q_{\rm c}=1, q_{\rm v} = 0) + \frac{3}{4}\epsilon_{\rm v}(q_{\rm c}=\sfrac{1}{3}, q_{\rm v} = 0)\right] & - \\
 \left[\frac{1}{4}\epsilon_{\rm c}(q_{\rm c}=1, q_{\rm v} = 0) + \frac{3}{4}\epsilon_{\rm c}(q_{\rm c}=\sfrac{1}{3}, q_{\rm v} = 0)\right]  \quad . &\label{eqn:gtp}
\end{align}
The Full Core Hole (FCH)\cite{Tanaka1999,Elsasser2001,Hetenyi2004} approach, in turn, excites a full core electron as in $\Delta$SCF \begin{align}
\nonumber \Delta E_{\rm FCH} = & \epsilon_{\rm v}(q_{\rm c}=0, q_{\rm v} = 0) - \\
		 	   & \epsilon_{\rm c}(q_{\rm c}=0, q_{\rm v} = 0) \quad . \label{eqn:fch}
\end{align}
The FCH approach was successfully applied to the simulation of X-ray absorption spectra of water and ice \cite{Hetenyi2004}, and fullerenes\cite{Nyberg1999,Deng2014}, while the TP approximation was found to perform well for organic molecules \cite{Triguero1998,Puttner2004,Diller2012}.
Interestingly, the obvious GTP analog to the GTS variant has not been considered before, and we include it in this study for completeness. 

The big numerical advantage of these implicit approaches is that a full NEXAFS spectrum can be obtained from a single (core-level constrained) DFT calculation, simply evaluating the transition energies to the different virtual KS states.
A certain disadvantage, especially with respect to an envisioned application to surface-adsorption systems typically calculated in periodic boundary condition supercells, is that an effectively charged system is created by removing (parts of) a core electron without compensating for it through the occupation of a virtual state. As such, the XCH approach \cite{Prendergast2006} is finally of particular interest. This approach creates a charge neutral final state by following the $\Delta$SCF philosophy to excite a full core electron to a virtual KS state. Simultaneously, however, it maintains the advantages of implicit variants by simply choosing the lowest unoccupied molecular orbital (LUMO) as this virtual KS state throughout. In other words, one occupation-constrained calculation is performed with the excited electron in the LUMO (designated by variable $q_{\rm l}$), and the entire spectrum is determined from it by reading off all virtual KS level positions \begin{align}
\nonumber \Delta E_{\rm XCH} = & \epsilon_{\rm v}(q_{\rm c}=0, q_{\rm l} = 1) - \\
		 	   & \epsilon_{\rm c}(q_{\rm c}=0, q_{\rm l} = 1) \quad .\label{eqn:xch}
\end{align}
Interestingly, the obvious transfer of this idea to the half core-hole TP and GTP approaches has also not yet been tried. To arrive at a systematic assessment, we therefore also consider corresponding XTP and XGTP occupation constraints in this study and will refer to this class of variants (XTP, XGTP, XCH) as charge-neutral implicit approaches, in contrast to the prior class of ionized implicit variants (TP, GTP, FCH) \begin{align}
\nonumber \Delta E_{\rm XTP} = \epsilon_{\rm v}(q_{\rm c} = 0.5, q_{\rm l} =
0.5) - \\
\epsilon_{\rm c}(q_{\rm c} = 0.5, q_{\rm l} = 0.5) \quad . \label{eqn:xtp} 
\end{align}
\begin{align}
	\nonumber \Delta E_{\rm XGTP} = \\
\nonumber\left[\frac{1}{4}\epsilon_{\rm v}(q_{\rm c}=1, q_{\rm l} = 0) + \frac{3}{4}\epsilon_{\rm v}(q_{\rm c}=\sfrac{1}{3}, q_{\rm l} =\sfrac{2}{3})\right] & - \\
\left[\frac{1}{4}\epsilon_{\rm c}(q_{\rm c}=1, q_{\rm l} = 0) +
\frac{3}{4}\epsilon_{\rm c}(q_{\rm c}=\sfrac{1}{3}, q_{\rm l}
=\sfrac{2}{3})\right] \quad .\label{eqn:xgtp}
\end{align}

\subsection{Preventing variational collapse}

The targeted non-ground-state KS occupation is the key conceptual aspect that distinguishes the various core-hole constraining variants. The major practical concern common to all variants is to achieve this occupation in the ensuing SCF cycle and prevent the variational collapse to the ground state. The objective is thus to identify in every SCF step of the constrained-occupation calculation which core orbital has the largest overlap with the targeted core orbital of the ground-state calculation so as to be able to enforce its occupation according to the recipe of the particular variant, cf. Table I. For the explicit approaches, the same holds for the identification of the virtual orbital that is to be filled, while for the charge-neutral implicit approaches, this holds for the identification of the LUMO. Recent approaches to this problem include local SCF (LSCF)\cite{Ferre2002,Loos2007}, linear expansion (le$\Delta$SCF)\cite{Gavnholt2008,Maurer2013}, constricted variational (CV-$\Delta$SCF)\cite{Ziegler2009,Park2015,Park2016,Cullen2011,Krykunov2013}, orthogonality constrained (OC-$\Delta$SCF)\cite{Evangelista2013,Derricotte2015} and $\sigma$-SCF\cite{Ye2017}. For the small molecular systems considered in this work, we instead maintain an originally specified occupational constraint during the SCF cycle by employing the maximum overlap method (MOM) \cite{Gilbert2008,Besley2009}. At every SCF step, this method evaluates which KS state has the largest overlap with the occupation-constraint KS state in the previous SCF step  and then modifies its occupation accordingly. To this end, it forms the orbital overlap matrix $\mathbf{O}$ \begin{equation}
	\mathbf{O} =
	(\mathbf{C}^{\textrm{old}})^{\dagger}\mathbf{S}\mathbf{C}^{\textrm{new}} \quad ,
\end{equation}
where $\mathbf{C}^{\textrm{old}}$ and $\mathbf{C}^{\textrm{new}}$ are the molecular-orbital coefficient matrices of the previous and current SCF iteration, respectively, and $\mathbf{S}$ is the overlap matrix. The projection of a state in the new KS eigenspace on the old eigenspace can then be written as \begin{equation}
	p_s = \sum_r^n O_{rs} = \sum_{\rm \nu}^{N}\left[\sum_{\mu}^{N}\left(\sum_{r}^{n}
	C_{r\mu}^{\textrm{old}}\right)S_{\mu\nu}\right]C_{\nu s}^{\textrm{new}} \quad .
\end{equation}
Here, $p_s$ is the projection of state $s$ in the subspace of the new KS eigenvector projected on the KS eigenvector in the previous iteration. $n$ spans all occupied states of the old KS eigenvector, and $\mu$ and $\nu$ are iterators over all basis functions of total number $N$. In our study of core-hole excitations, we want to propagate a single state through the SCF cycle. Therefore, we project the previously constrained KS state on to a subspace of the new eigenvector to identify the new state with a modified occupation, essentially inverting the typical MOM-procedure. To this extent we calculate the projection $\tilde p_s$ of the constrained state on a subspace spanning from $n_{\rm i}$ to $n_{\rm f}$. The occupational constraint is then propagated on the KS state of largest $\tilde p_s$ \begin{align}
	\tilde p_s = \sum_{\nu}^N \left[ \sum_{\mu}^N \left(
			\sum_{r=n_{\rm i}}^{n_{\rm f}}
		C_{r\mu}^{\textrm{new}}\right)S_{\mu\nu}\right]C_{s\nu}^{\textrm{old}} \quad . \label{eqn:inverted_mom}
\end{align}

\begin{figure}
	\includegraphics[width=0.45\textwidth]{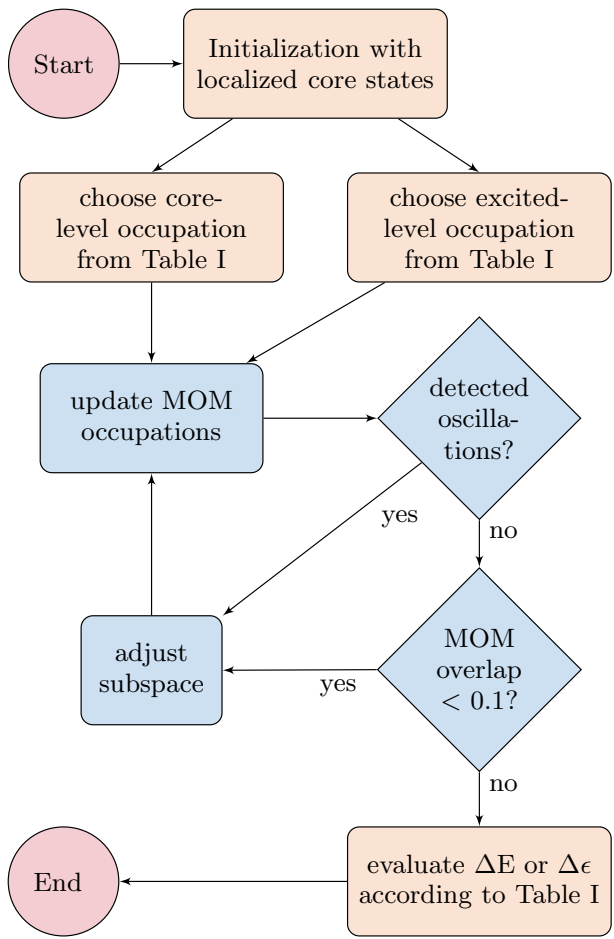} 
	\caption{Employed workflow to achieve robust NEXAFS simulations using the maximum-overlap method (MOM). The part highlighted in blue is executed at each SCF step until SCF convergence is achieved.}
	\label{fig:workflow}
\end{figure}

In the original MOM approach the subspace to be projected on was split into all occupied and all virtual KS states of the ground state calculation, and then constraints to enforce the hole and to enforce the occupation of a virtual state were separately projected on each manifold. In this work, we found this most general procedure to lead to massive convergence problems (SCF oscillations). We therefore developed a more restricted approach as follows: For K-edge NEXAFS, we are specifically interested in the lowest-energy 1s states of carbon (or nitrogen). We therefore restrict the occupied subspace to the $m$ degenerate lowest-energy KS states of the ground-state calculations for a molecule containing $m$ C (or N) species. For those variants that additionally require an enforced occupation of an virtual state LUMO+$k$, we initially define the unoccupied MOM subspace to only consist of the ground-state orbitals {[LUMO, LUMO+$k$]}. This considers that the occupation of a virtual orbital typically lowers its energy. We found that only in a few cases, state reordering shifts the targeted KS state above this range. In those cases, reflected by MOM overlaps ($\tilde p_{s}$) below 10\% 
we then gradually expanded the MOM subspace to {[LUMO, LUMO$+k+x$]}, $x>1$ until higher overlaps where found. In those cases, where oscillations between degenerate orbitals still prevail in the restricted {[LUMO, LUMO+$k$]} subspace, we instead gradually shrank the subspace further to {[LUMO, LUMO$+k-x$]}, $x>1$.
A schematic workflow of our approach is shown in Fig.~\ref{fig:workflow}. We validated that this workflow led to the correct occupations by comparing the initial and final eigenvector belonging to the state with a modified occupation in terms of their major constituent basis functions. We find that the principal character of the KS state does not change if we apply our modified MOM-procedure. A comparison of our modified MOM procedure in comparison with the original approach including occupational smearing is provided in the supporting information. While the modified MOM-procedure thus enabled the systematic benchmark performed in this work, we nevertheless emphasize that reaching convergence and correct occupations in case of the explicit variants is a strenuous endeavor that requires a lot of human interference and control, as is also highlighted in the supporting information where our modified method, although prevailing over the original approach, can not resolve the entirety of explicitly occupied virtual states. This is another aspect that strongly favors the implicit variants, for which achieving correct occupations of the modified core state was generally found to be straightforward with our modified MOM-procedure.
We note in this respect, that a popular alternative to the MOM method in plane-wave implementations of DFT is the usage of pseudopotentials, where either the atom carrying the core-hole is described by a pseudopotential created with a core-ionization\cite{Gao2009,Diller2017,Su2017,Prendergast2006} or through the reverse strategy of self-consistently determining the core-hole state (described in an all-electron form) and replacing all other atoms of the same species by an effective pseudo-potential (ECP)\cite{Triguero1998,Minkov2004,Kolczewski2001}.
Either way, the requirement for a mechanism to keep the core-hole localized is lifted.

\subsection{Computational Details}

The collinear spin-resolved DFT calculations were performed using the FHI-aims package \cite{Blum2009,Ren2012}. Electronic exchange and correlation (xc) were treated on the generalized-gradient approximation (GGA) level with the PBE\cite{Perdew1996} functional and at the hybrid functional level with the PBE0\cite{Adamo1999} functional. The ground-state geometry of all molecules was fully relaxed until residual forces were below 10$^{-3}$eV/{\AA}. The occupational-constraint excited state calculations were then conducted on these optimized ground-state geometries. 

FHI-aims uses numeric-atomic orbital (NAO) localized basis sets. The standard basis sets for semi-local functionals are categorized into tier levels of increasing basis set size and accuracy. Basis set convergence was evaluated by running tier1, tier2, and tier3 calculations. As further detailed in the original FHI-aims publication, \cite{Blum2009} the tier1 set consists of the minimal basis (chosen as the solution of the free atom) and, additionally, ionic and hydrogenic basis functions, determined in an automated procedure and ordered by their magnitude of improvement of interatomic binding energies. The tiers naturally arise as groups of different angular momenta \textit{spd} (tier1), \textit{spd(f,g)} (tier2), etc. similar to the intuitive construction in Gaussian basis sets \cite{Wilson1996,Frisch1984,Weigend2005a} and are hierarchically organized, with a higher tier always including all functions of the lower tier. The tier basis sets were constructed and optimized for total energy differences and the usage at the local-density (LDA) and generalized-gradient approximation (GGA) functional level.\cite{Blum2009} They may be used for higher-rung functionals, too. However, a valence-correlation-consistent NAO-VCC basis set family \cite{Zhang2013} has been specifically constructed for such calculations, following the same principle as also used for Dunning-type Gaussian basis sets \cite{Dunning1989} at the cc-pV2Z, cc-pV3Z, and cc-pV4Z level. In terms of available basis functions, these basis sets are comparable to the tier1, tier2, and tier3 basis sets, respectively. For all basis sets, integration on the numerical grids was carried out at the "tight" level implemented in FHI-aims.\cite{Blum2009}

For the occupation-constraint calculations, the core-state orbitals were first maximally localized at the end of the ground-state calculation by following the procedure outlined by Foster and Boys\cite{Foster1960,Rossi2016}. This was then used as an initial guess with the modified occupations as shown in Figure \ref{fig:approximations} and preventing variational collapse during the ensuing SCF cycle following the MOM scheme described in Fig. \ref{fig:workflow}. To determine intensities belonging to each transition energy, we evaluated the transition dipole moment between the core state i and each unoccupied state f entering eq.~(\ref{eq1}) as \begin{align}
	{\bf \mu}_{\rm if} = \braket{\phi_{\mathrm{i}}|\hat x|\phi_{\mathrm{f}}}\quad ,
\end{align}
where $\phi_{\mathrm{i}}$ and $\phi_{\mathrm{f}}$ are the KS eigenvectors of states i and f, respectively. For the explicit models, each transition dipole moment was determined from the corresponding calculation with a modified final state occupation of state f. 

\section{Results}

\subsection{Benchmark approach}

\begin{figure}[htbp!]
	\includegraphics[width=0.45\textwidth]{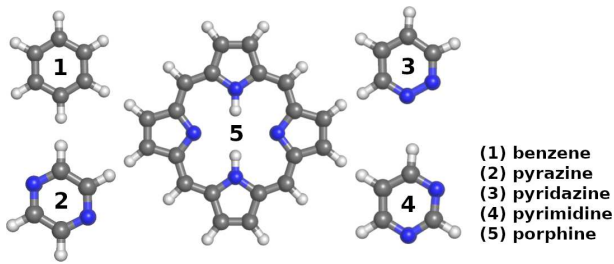}
	\caption{C and N containing molecules forming the considered benchmark set (C = gray spheres, N = blue spheres, H = white 
	spheres).}
	\label{fig:ball_and_stick}
\end{figure}

For our benchmark study we consider the five molecules shown in Fig.
\ref{fig:ball_and_stick}: benzene (C$_6$H$_6$) and the four heterocyclic
molecules pyrazine (C$_4$N$_2$H$_4$), pyridazine (C$_4$N$_2$H$_4$), pyrimidine (C$_4$N$_2$H$_4$), and porphine (C$_{20}$N$_4$H$_{14}$). The small size of the molecules and the $C_2$ rotational axis present in their gas-phase structure would in principle readily allow for highly accurate computational spectroscopy approaches. However, when adsorbed at a transition metal surface, the likely break of symmetry\cite{Fronzoni2012} and the necessity to explicitly treat the extended surface in a periodic boundary supercell\cite{Diller2017} approach rapidly increases the computational cost to render effective core-hole constraining approaches an appealing option. We compare the calculated NEXAFS spectra to experimental data either from gas-phase measurements (benzene\cite{Puttner2004}, pyridazine\cite{Vall-llosera2008}, pyrimidine\cite{Vall-llosera2008}, pyrazine\cite{Vall-llosera2008}) or from multilayer magic-angle measurements, where no angle-dependency is present and the molecule-surface interaction can be neglected (porphine\cite{Bischoff2013}). In our comparison we specifically focus on the near-edge region and therefore consider the three lowest-energy excitations/peaks. This corresponds to an interval of approximately 3\,eV above the carbon 1s edge and approximately 6.5\,eV above the nitrogen 1s edge. To quantify the deviation from the experimental signatures, we measure the error in the simulated peak position relative to the first edge peak \begin{equation}
	\textrm{error}_{\rm eng} [\%] = 100 - \frac{\textrm{(peak energy - edge
			peak energy)$_{\rm comp}$}}{\textrm{(peak energy - edge 
	peak energy)$_{\rm exp}$}} \quad . \label{def_eng_error}
\end{equation}
This measure of the error in energy is taken relative to the correct (experimental) value and both over- and under-estimation of the excitation energy is captured in the following analysis, where a positive value corresponds to an underestimation and a negative value to an overestimation of the transition energy. An equivalent approach is pursued for the simulated intensities, here normalizing to the edge peak intensity \begin{equation}
	\textrm{error}_{\rm int} [\%] = \frac{\textrm{(peak intensity/edge peak
			intensity)$_{\rm comp}$}}{\textrm{(peak intensity/edge 
	peak intensity)$_{\rm exp}$}} \quad .
	\label{def_int_error}
\end{equation}
For the considered molecules, the experimental near-edge spectrum corresponds primarily of well-separated high-intensity peaks. This allows for a facile identification and assignment of the peaks. Only in a few cases, particularly for the C-edge spectra of the larger compound porphine, some experimentally observed peaks are made up from two (or more) overlapping resonances. In this case, we used the higher intensity resonance for the benchmarking and are well aware of the possible (small) systematic error thus included in our analysis.

\subsection{Method comparison: Transition Energies}

\begin{figure}[htbp!]
	\includegraphics[width=0.45\textwidth]{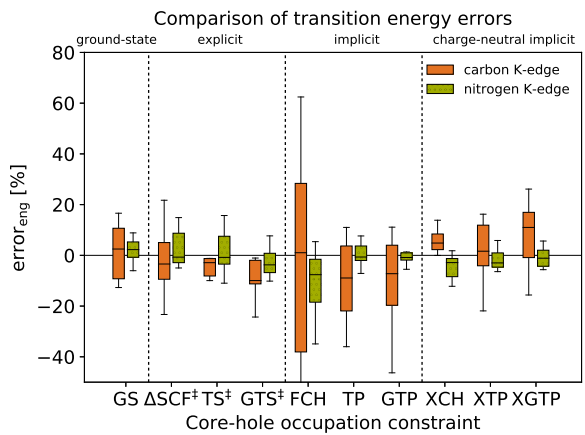}
	\caption{Box plot of the error in the transition energies as defined in eq.~(\ref{def_eng_error}) for the considered range of core-hole occupation constraining approaches (using DFT-PBE and a tier2 basis set). The upper and lower limits of the 
	rectangles (interquartile range, IQR) mark the 75\% and 25\% percentiles, the internal horizontal line marks the median, and the "error 
	bars" mark the 99\% and 1\% percentiles (defining the maximum absolute errors, MAEs).}
	\label{fig:bar_energy}
\end{figure}

Figure \ref{fig:bar_energy} compiles the box-plots of the error in the calculated transition energies for the different core-hole occupation constraining approaches. Here we first focus on the PBE functional and tier2 basis sets, as this would currently correspond to the affordable state-of-the-art to describe metal-adsorption systems (or possibly including a +U correction for semiconductor-adsorption systems). We return to a discussion of the xc functional and the basis set size dependence below. The median error of all approaches is rather low and lies generally around and below 10\%. A
notable exception are the transition potential based approaches (TP, GTP and XGTP), which seem to have a particular problem with reproducing the carbon peak positions. Among the computationally most demanding explicit variants, the TS approach performs best (median error: -2.9\%). However, the best performing
charge-neutral implicit approaches, XTP (median error: +1.6\%) and XCH (median error: +4.8\%)
are comparably good. Within the considered near-edge region, peak position deviations around 10\% correspond to absolute errors below a few tenths of eV,
in line with 1s$\rightarrow \mathrm{\pi}^{*}$ excitation energy accuracies of previous reports in the GTS approximation\cite{Hu1996,Chong1996,Triguero1999}.
This would generally be sufficient for an assignment of experimental spectra as exemplified below. The superficial look at the median error would therefore suggest essentially all of the tested variants as viable. 

A more differentiated view is instead obtained from the more detailed analysis of the interquartile range (IQR) and maximum absolute errors (MAEs) also contained in the box plot in Fig.~\ref{fig:bar_energy}. Here, clear performance differences arise between the different variants, revealing partly exceedingly large errors. In particular, for the FCH variant, the low median error seems to arise from a favorable cancellation of partly unacceptably large errors.
Reports, which prefer this FCH variant over other approximations do this on account of a better description of the intensities\cite{Hetenyi2004,Nyberg1999}, which, as discussed in a moment, is indeed the case. Other authors also report exceedingly large energetic deviations of the FCH variant, with much better results obtained from a TP\cite{Triguero1998}, or even a GS calculation\cite{Laskowski2009}.
 Interestingly, also with respect to the IQR, which contains 50\% of the data and thus spans from the lower to the upper quartile, and the MAEs, there is a  significant element-specific performance, with all variants better able to reproduce the nitrogen spectra. 

Henceforth considering the IQR as a good performance indicator, we also arrive at partly unexpected insights regarding the approximation of the excited-state energy itself. The consideration of the core-hole relaxation energy contribution through the explicit change of level occupations in the $\Delta$SCF method does intriguingly not lead to a dramatic performance improvement as compared to a straightforward ground-state calculation. In fact, in case of the nitrogen 1s peaks, it even worsens the IQR. Even more surprisingly, the TS and GTS approaches, which are in principle nothing but a reformulation of the $\Delta$SCF approach plus an integral approximation, lead to somewhat improved IQRs as compared to $\Delta$SCF itself. This suggests a favorable cancellation of errors either within these effective approaches or in the interplay with the approximate DFT functional. Such cancellation effects would also help to understand why the complete neglect of the excited electron in the implicit TP and GTP variants apparently lowers the N 1s IQR compared to the physically more accurate explicit approaches, whereas the C 1s IQRs show the expected trend with explicit variants exhibiting the lowest IQR, implicit variants (FCH, TP, GTP) the highest IQR -- and the implicit charge-neutral variants (XCH, XTP, XGTP) with their average consideration of the excited electron somewhat performing intermediate between these two.

\begin{figure}[htbp!]
	\includegraphics[width=0.45\textwidth]{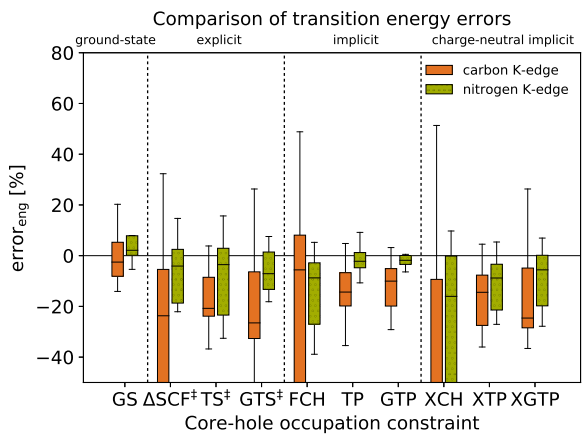}
	\caption{Same as Fig.~\ref{fig:bar_energy}, but now using the hybrid functional PBE0 and a NAO-VCC-3Z basis set.}
	\label{fig:bar_energy_hybrid}
\end{figure}

In order to assess the role of the DFT functional in such error cancellation we, therefore, repeated all calculations with the hybrid functional PBE0. For the pure gas-phase molecules, this functional will definitely yield a significantly improved ground-state electronic structure. The results obtained with the FHI-aims NAO-VCC-3Z basis set are summarized in Fig.~\ref{fig:bar_energy_hybrid}.
While no experience with this recommended basis set class for levels of theory including exact exchange exists for NEXAFS calculations, triple-zeta type Gaussian bases are frequently recommended for the calculation of core ionization or core excitation \cite{Tolbatov2014,Takahata2000,Takahata2003,Chong2002}.
Intriguingly, we obtain a rather mixed result. For a few variants (GS, TP, GTP) we obtain the anticipated improvement with this higher-rung functional, in particular with respect to the C1s IQRs that were found to be particularly problematic at the DFT-GGA level. For all others, errors, in fact, increase at least by 10-20\%. A closer look reveals that these are unanimously those
variants that include the occupation of virtual states, i.e., the explicit variants ($\Delta$SCF, TS, GTS) and the charge-neutral implicit variants (XCH, XTP, XGTP). The FCH approach remains in its IQR performance as abysmal as it was before. 

We should note that the performance of a hybrid functional for the virtual KS states of the electronic ground state was for instance already analyzed previously by van Meer {\em et al.} (there as basis for a time-dependent DFT-treatment)\cite{VanMeer2014}. The conclusion was that the eigenstates were too diffuse, in some instances even unphysical. Other authors also reported large error bars in $\Delta$SCF calculations of 0.5~eV for first-row elements and 1.5~eV for second-row elements when using a hybrid functional\cite{Besley2009}.
This could suggest that the partly good performance obtained for these variants at the PBE level results indeed (largely) from an effective error cancellation between approximate semi-local DFT functional and effective treatment of the core-excitation. To this end, we also have to note the construction concept of the FHI-aims NAO basis sets though. While the correlation-consistent Dunning Gaussian basis sets were validated based on single and double excitations\cite{Dunning1989}, the basis functions in both classes of FHI-aims basis sets (tier and NAO-VCC-nZ bases) targeted the total ground-state energy. In NAO-VCC-nZ this relates to spherically symmetric atoms in the frozen-core random-phase approximation\cite{Zhang2013}, and for the tier basis sets, the total energy of atomic dimers in the LDA-approximation \cite{Blum2009}. They were hitherto only validated to perform well for covalent bonds and isomerization energies. In particular, the lack of additional diffuse functions (present in the aug-type Dunning basis sets) and the optimization of valence-correlation consistency only in the NAO-VCC-nZ bases, could yield a particularly bad description of the energy differences between occupation-constrained core and valence states entering the NEXAFS transition energies. The really slow basis set convergence described below for the NAO-VCC-nZ bases could indeed hint at the inadequacy of these basis sets in describing these important KS energy differences. 

While a full identification of the reason behind the poor performance of explicit and implicit charge-neutral variants at hybrid level has to await the construction of new tailored basis sets (which is beyond the scope of the present study), we note that it is predominantly GGA functionals that are currently of interest/affordable for surface-adsorption calculations. In fact, already the PBE0/NAO-VCC-3Z calculations behind the gas-phase molecule benchmark of the explicit variants in Fig.~\ref{fig:bar_energy_hybrid}
involved a computational cost that fully defeats the purpose of these effective NEXAFS simulation approaches.

\subsection{Method comparison: Transition Intensities}

\begin{figure}[htbp!]
	\includegraphics[width=0.45\textwidth]{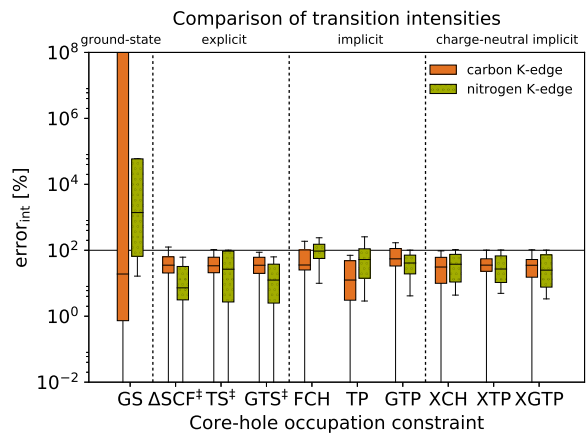}
	\caption{Same as Fig.~\ref{fig:bar_energy} (using DFT-PBE and a tier2 basis set), but now for the error in the transition intensities.}
	\label{fig:ibar_minmax}
\end{figure}

For computational spectroscopic support, a reliable description of the peak intensities is almost as important as the correct description of the peak energies. In Fig.~\ref{fig:ibar_minmax} we, therefore, compile the determined errors in the intensities as evaluated according to eq.~(\ref{def_int_error}), and again explicitly summarizing mean errors, IQRs, and MAEs in the shown box plot. Here, a simple ground-state calculation is clearly inadequate, with exceeding errors in all three performance indicators. All other variants perform significantly better, in fact with not too much variation between them. Their median is consistently below 100\%, which means that the intensity of the edge
peak is consistently overestimated with respect to the other higher transitions.
We suspect additional contributions to the experimental intensity of the edge peak as a possible reason for this consistent overestimation. One contribution could come from forbidden transitions, which receive finite intensity in experiment through thermally induced motion/symmetry breaking and thus reduce the actual intensity of the edge peak.
There are currently two major ways as of how these broadening of the peaks and the associated decrease in main peak intensity can be included in the simulation, either by resolving the vibronic structure (coupling of vibrational and electronic states via the linear coupling model\cite{Domcke1977,Plashkevych2000,Kolczewski2001,Fronzoni2014}) or by following the Herzberg-Teller effect\cite{Herzberg1933} and including temperature broadening by either doing classical or ab-initio MD sampling of the system and then averaging over different snapshots of the trajectory\cite{Jahn2014,Hetenyi2004,Chen2010,Iannuzzi2008,Wernet2004,Kong2012,Schwartz2009,Schwartz2009a,Schwartz2010,Matsui2016}.
Another methodology calculates the spectroscopic signatures for geometries at the turning point of each vibrational mode\cite{Perera2018}.
%

\begin{figure}[htbp!]
	\includegraphics[width=0.45\textwidth]{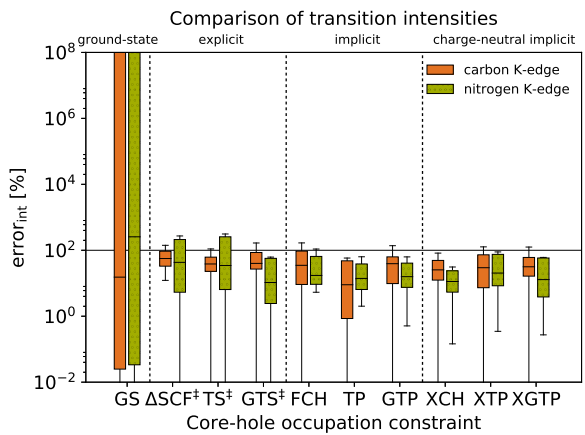}
	\caption{Same as Fig.~\ref{fig:ibar_minmax}, but now for the error in
	the transition intensities using PBE0 and a NAO-VCC-3Z basis set.}
	\label{fig:ibar_minmax_hybrid}
\end{figure}

\begin{figure}[htbp!]
	\includegraphics[width=0.45\textwidth]{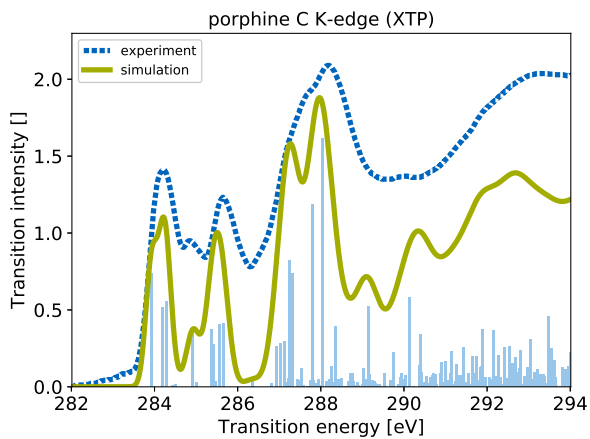}
	\caption{Experimental\cite{Bischoff2013} and simulated NEXAFS carbon K-edge spectrum of the
		porphine molecule. Simulation at the PBE/tier-2 level of
		theory using the XTP approximation. Shown are the calculated transition energies (light
		blue) and a spectrum generated by the superposition of the Gaussian-broadened delta peaks
		with linearly increasing broadening between 0.32 and 1.6 eV
		toward higher transition energies, which is common practice in
		the computational analysis of NEXAFS spectra\cite{Kolczewski2001,Kolczewski2003}
		(green curve).}
	\label{fig:porphine_xgtp}
\end{figure}

While such vibrational simulations would certainly be desirable, we note that apart from this overestimation of the edge peak all IQRs are consistently small.
This demonstrates that even without such vibrational corrections, essentially all variants will be able to reliably determine the remaining spectral profile. Noteworthy, the FCH approach has the smallest IQR and is the only variant with an IQR partially above 100\%, as had also been noticed
for GGA-type calculations of water\cite{Prendergast2006}. This holds as well for our benchmarks at the hybrid functional PBE0 level, which we compile for completeness in Fig.~\ref{fig:ibar_minmax_hybrid}, even though as discussed above there are presently clear issues with calculations at this level of theory in FHI-aims. Correspondingly, we also exemplify the reliable determination of the spectral profile for the GGA-level aspired for the surface-adsorption context, and in particular in Fig.~\ref{fig:porphine_xgtp} we show a comparison of experimental data for the porphine molecule \cite{Bischoff2013} and the constituent resonances as determined by the simulation using the charge-neutral implicit XTP variant.

\subsection{Basis set dependence}\label{sec:basis}

\begin{figure}[t!]
	\includegraphics[width=0.45\textwidth]{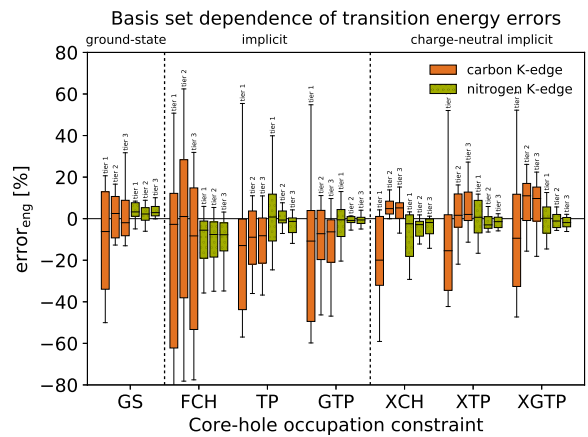}
	\includegraphics[width=0.45\textwidth]{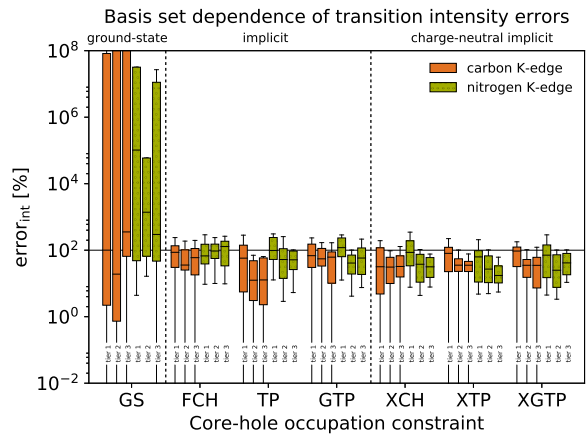}
	\caption{Convergence of the excitation energy (upper panel) and excitation intensities (lower panel) with respect to the hierarchical tier basis sets for PBE calculations. See Fig. \ref{fig:bar_energy} for an explanation of the shown box plots.}\label{fig:convergence_pbe}
\end{figure}

\begin{figure}[t!]
	\includegraphics[width=0.45\textwidth]{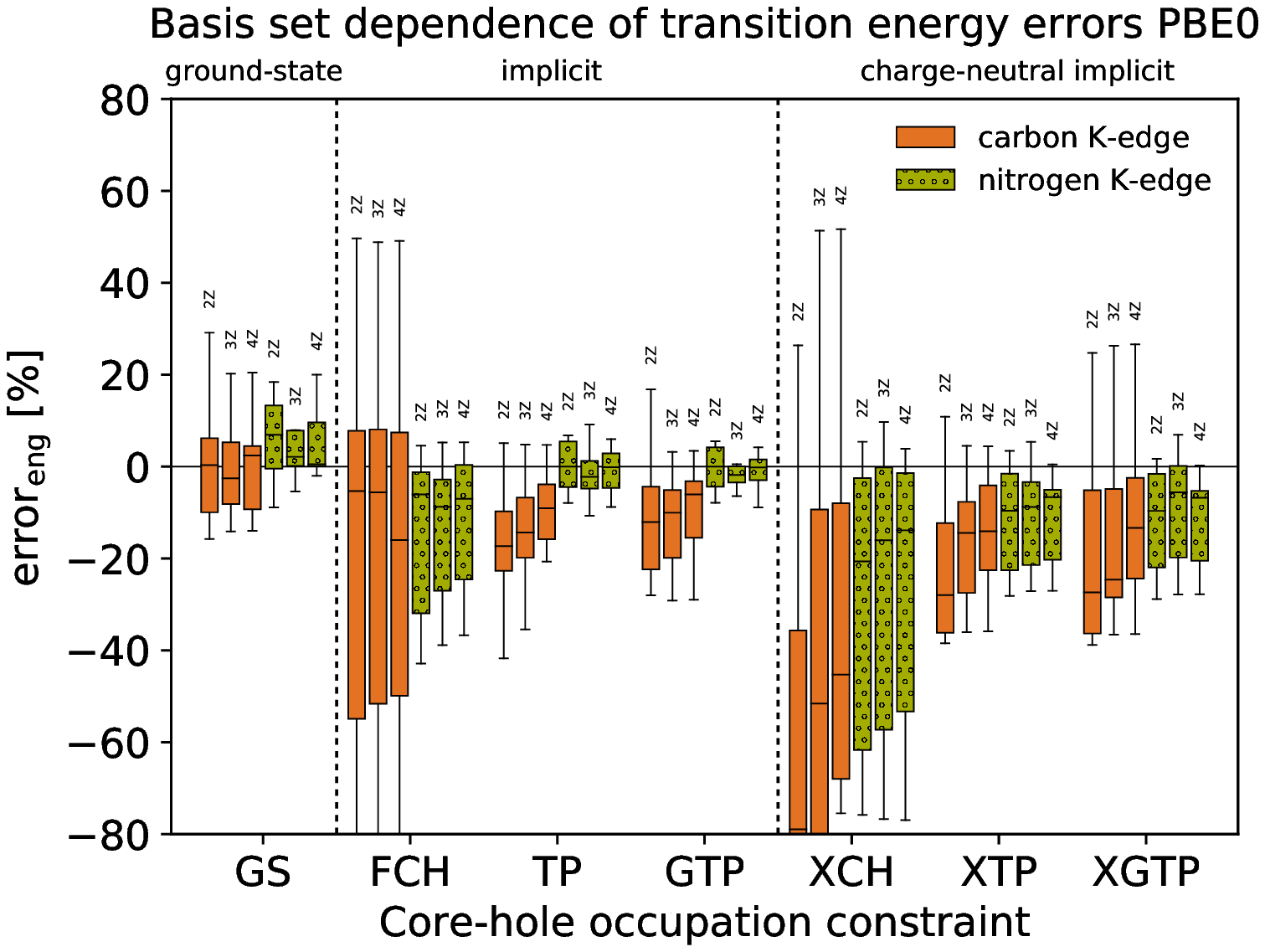}
	\includegraphics[width=0.45\textwidth]{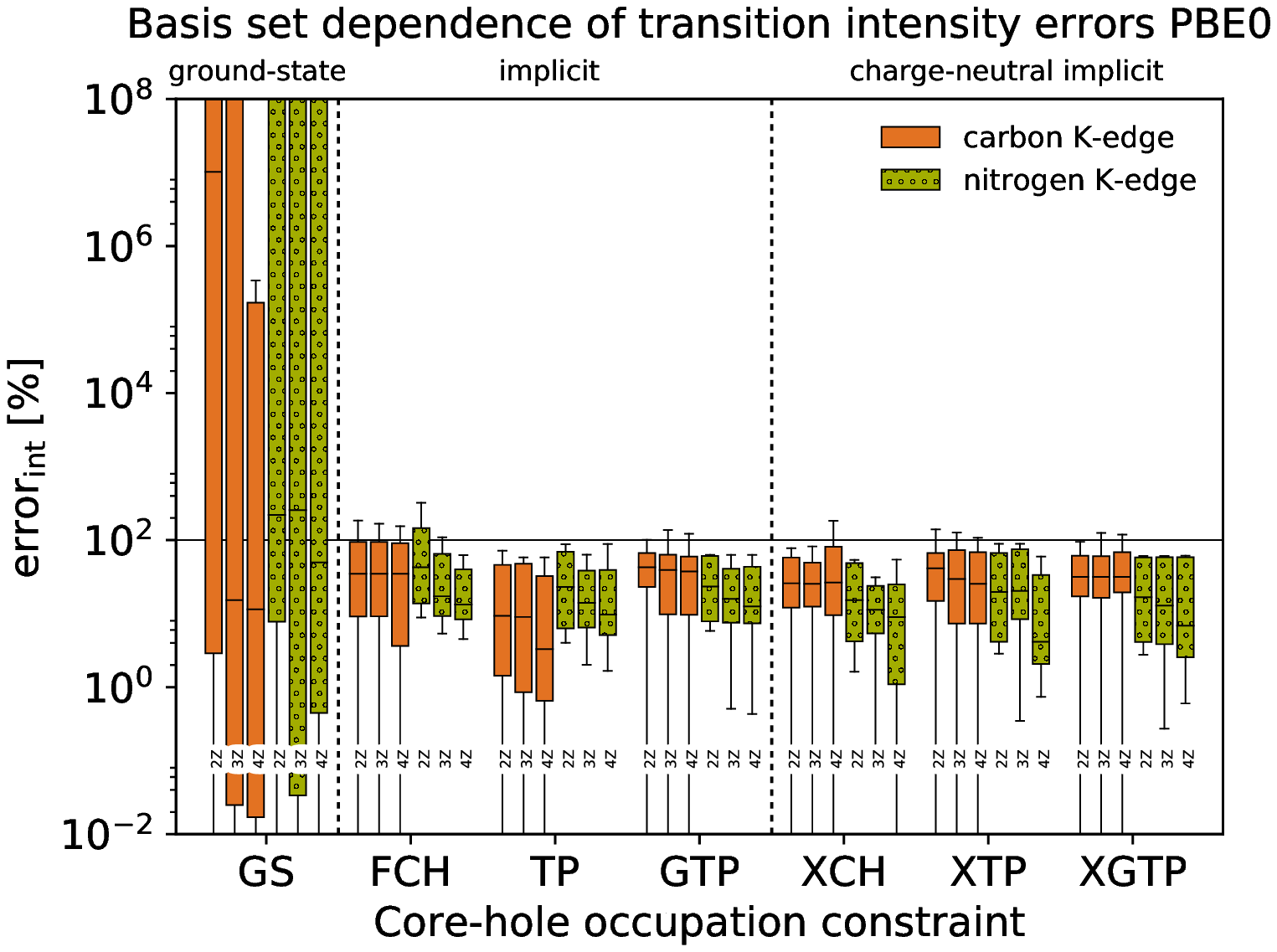}
	\caption{Same as Fig.~\ref{fig:convergence_pbe}, but for the hierarchical NAO-VCC-nZ basis sets for PBE0 calculations.}\label{fig:convergence_pbe0}
\end{figure}

NEXAFS probes the unoccupied, more delocalized states of the given system. One would therefore generally expect a slower convergence with basis set size for localized bases\cite{Boffi2016}, even if only the energetically lowest-lying unoccupied states are targeted in simulations of the near-edge region. This has been confirmed in the spectroscopic context in simulations of excitations to outer-shell valence states using the $\Delta$SCF method\cite{Chong2006}. As shown in Fig.~\ref{fig:convergence_pbe} for PBE and in Fig.~\ref{fig:convergence_pbe0} for PBE0 we indeed observe consistent improvements in the transition energies notably for the IQRs when increasing the hierarchic tier and NAO-VCC-nZ basis sets in FHI-aims. Nevertheless, at the tier2 and NAO-VCC-3Z basis sets employed as default in the previous sections a convergence is reached that justifies the conclusions made. We expect a similar convergence also for the explicit variants for which we could not afford a systematic convergence test at the largest (tier3, NAO-VCC-3Z) basis set, partly due to insuperable convergence problems.

At the PBE level, the smallest tier1 basis set is clearly not apt to describe the transitions, and we observe partly abrupt changes to the next larger tier2 basis set, in particular in the more sensitive performance indicators IQR and MAE. Here, we ascribe the especially pronounced improvements in the carbon spectra for instance to the additionally available p-type basis functions in the tier2 set. Further available functions in the tier3 set do not seem to lead to any systematic improvement, but this might also simply be masked by error cancellation with the approximate DFT functional as discussed above. Literature is also not clear at this point, with diffuse functions once found to be required in GGA calculations of K-edge absorption spectra of small molecules using $\Delta$SCF \cite{Chong2006}. In contrast and also at the GGA level, van Meer {\em et al.} report that a large basis set introduced a clustering of many spurious states with low oscillator strengths in the virtual space at an energy of -$\epsilon_{\mathrm{HOMO}}$, above which the states do not correspond well to excitation energies anymore\cite{VanMeer2014}. Similarly, when using hybrid DFT functionals in another study \cite{Fouda2018}, the authors also experienced that including too many diffuse basis functions can lead to a decrease in accuracy -- a behavior we also observe for many variants in the nitrogen transition energies. 

Generally, however, we emphasize particularly the different convergence behavior of the different variants and of the MAEs at the two functional levels. In our view, the prior clearly indicates again quite a degree of unsystematic error cancellation between finite basis set and effective treatment of the excited state energy. In turn, the latter seems to support our assessment that there is a general problem with the presently available NAO-VCC-nZ basis sets for such kind of simulations. In fact, we obtain even worse performance and comparably bad convergence behavior when using the tier basis sets designed for the semi-local functionals in the PBE0 calculations. As already seen when comparing the different variants in Section III.C the transition intensities are much less demanding in this respect. Satisfactory relative intensities will already be obtained with moderate basis sets and quite consistently over all variants.

\section{Conclusion and Outlook}

We have systematically assessed a wide range of variants within the core-level occupation constraining approach to simulating NEXAFS spectra. Its comparably high numerical efficiency makes this approach particularly appealing for (very) large systems as typically encountered in the context of supra-molecular assembly or surface adsorption. At the same time, its highly approximate treatment of the excited state energies calls for systematic tests concerning its reliability. Using a dedicated set of C- and N-containing molecules, our benchmark indeed points at quite some degree of error cancellation between the effective treatment of the excited state energy, the approximate exchange-correlation functional and the finite localized basis set used in the underlying DFT calculations. Focusing not only on the average reproduction of transition energies and intensities, but also considering more sensitive performance indicators like the interquartile range and maximum absolute error, our study identified in particular the presently available hierarchical numeric-atomic orbital basis sets (tier and NAO-VCC-nZ) in the general program package FHI-aims as not suitable for NEXAFS simulations on the basis of higher-rung functionals including exact exchange.

For the representative semi-local DFT functional PBE, we instead find all tested variants to overall yield reliable spectra already at a moderate basis set size.
Reliable here refers to an accuracy that affords a semi-quantitative analysis of experimental data. Particularly appealing for surface adsorption calculations are the so-called charge-neutral implicit variants, as they conform easily with periodic boundary condition supercells. Within this class of variants, we find in particular the XCH and XTP variants to perform most robustly in our benchmark, with the latter variant in our view having a somewhat better motivated mathematical basis (and thus potentially exhibiting a better transferability).
The implicit nature of these variants, i.e. allowing to obtain a full NEXAFS spectrum out of one constrained-occupation DFT calculation, renders them numerically much more feasible than earlier explicit variants, for many of which in fact already the compilation of the NEXAFS spectra for the present set of gas-phase molecules becomes real cumbersome. Simultaneously, the here established protocol of preventing variational collapse of the exciton through the maximum overlap method ensured a swift self-consistency for these charge-neutral implicit variants as known from the alternative class of ionized implicit variants that completely neglect the excited electron in the unoccupied subspace.

\section*{Supplementary Material}
Supplementary material including more details on our modified maximum overlap
method (MOM) as well as the convergence behavior of explicit methods for small
molecules is available online.

\section*{Acknowledgments}
The authors gratefully acknowledge support by the TUM Institute for Advanced Studies and the compute and data resources provided by the Leibniz Supercomputing Centre (www.lrz.de).  We are also thankful for fruitful discussions with Katharina Diller, Reinhard Maurer, and Volker Blum.

\bibliography{jcp2016}

\end{document}